\documentclass[acmsmall,nonacm]{acmart}

\usepackage{booktabs}
\usepackage{amsmath}
\usepackage{array}
\usepackage{graphicx}
\usepackage{xcolor}
\newcolumntype{P}[1]{>{\raggedright\arraybackslash}p{#1}}
\acmJournal{TOSEM}
\acmYear{2026}
\acmMonth{7}
\setcopyright{none}
\newcommand{\DeclareResult}[2]{\expandafter\gdef\csname result@#1\endcsname{#2}}
\newcommand{\result}[1]{%
  \ifcsname result@#1\endcsname
    \csname result@#1\endcsname
  \else
    \textcolor{red}{\textbf{[[#1]]}}%
  \fi}
\DeclareResult{R-ABLATE-P3}{$-$47.2 percentage points}
\DeclareResult{R-ABLATE-P4}{$-$27.4 percentage points}
\DeclareResult{R-ALL-PAIR-AUDIT}{considered all 4,800 operator--parent pairs, all of which were applicable, activated, isolated, and linked to an unchanged parent digest; 1,440 were sampled and 3,360 remained unsampled}
\DeclareResult{R-CONCLUSION-CLAIM}{the calibrated claim that executable domain contracts expose encoded release violations beyond conventional tooling, not that the method estimates production defect prevalence or guarantees safe plugins}
\DeclareResult{R-CONCLUSION-EFFECT}{increased controlled-mutant recall from 25.3\% to 100\%, a 74.7 percentage-point difference (clustered 95\% CI 73.4--75.8), with 0/120 observed clean-package rejections}
\DeclareResult{R-CONCLUSION-GENERALIZATION}{preserved the constructed-stratum advantage on held-out specifications and the all-specification operator challenge, while a bounded-repair generated stratum supplied downstream evidence without replacing the zero-yield strict endpoint}
\DeclareResult{R-CONFIRM-DESIGN}{600 unmodified generations from two current coding models, two prompt conditions, five seeds, and 30 specifications}
\DeclareResult{R-CONFIRM-DEVSTRAL}{all 300 Devstral outputs failed the interface parser after being wrapped in an unlabelled outer Markdown fence}
\DeclareResult{R-CONFIRM-P0-FULL}{0/600 passed P0 and 0/600 achieved full P0--P4 conformance (each 95\% CI 0--0.64\%)}
\DeclareResult{R-CONFIRM-PARSE}{50.0\% (300/600; 95\% CI 46.01--53.99\%)}
\DeclareResult{R-CONFIRM-PROMPT}{0.0 percentage points for parse, P0, and full-conformance yield (specification-clustered 95\% intervals 0.0--0.0)}
\DeclareResult{R-CONFIRM-QWEN}{all 300 Qwen outputs parsed but none passed P0 because required visible manifest fields were missing}
\DeclareResult{R-CONFIRM-RAW}{600/600 generation records}
\DeclareResult{R-CONVENTIONAL-CLEAN-FPR}{0\% (0/120; 95\% CI 0--3.1\%)}
\DeclareResult{R-CONVENTIONAL-RECALL}{25.3\% (365/1,440; 95\% CI 23.2--27.7\%)}
\DeclareResult{R-DISCUSSION-LAYERED-EVIDENCE}{a 74.7 percentage-point recall increase in the constructed stratum with no observed increase in clean-package rejection}
\DeclareResult{R-DISCUSSION-LAYERED-IMPLICATION}{that executable domain obligations covered defect classes that build, security, and dependency checks did not address in this benchmark}
\DeclareResult{R-DISCUSSION-PROMPT-EVIDENCE}{the all-zero full-conformance outcome under both prompts and a 7.0-point lower format-normalized parse yield under the assurance-aware condition}
\DeclareResult{R-DISCUSSION-PROMPT-IMPLICATION}{that prompt elaboration did not overcome basic interface and functional failures in this setting}
\DeclareResult{R-EXTERNAL-DESIGN}{16 vulnerable/fixed announcement pairs after two predeclared exclusions}
\DeclareResult{R-EXTERNAL-DIRECTION}{directional finding reduction in 0/16 pairs}
\DeclareResult{R-EXTERNAL-FIXED}{0/16 fixed revisions}
\DeclareResult{R-EXTERNAL-MCNEMAR}{0 discordant pairs (exact McNemar $p=1.0$)}
\DeclareResult{R-EXTERNAL-RESULTS}{the frozen generic PHP detector found 0/16 vulnerable revisions (95\% CI 0--19.36\%), rejected 0/16 fixed revisions, and showed directional reduction in 0/16 pairs}
\DeclareResult{R-EXTERNAL-VULNERABLE}{0/16 vulnerable revisions (95\% CI 0--19.36\%)}
\DeclareResult{R-FACTORIAL-CHALLENGE}{P0--P2 recall of 25.0\% (240/960) and P0--P4 recall of 100\% (960/960), a 75.0-percentage-point difference (fixed-set design-resampling range 75.0--75.0)}
\DeclareResult{R-FACTORIAL-DIFF}{75.0-percentage-point}
\DeclareResult{R-FULL-CLEAN-FPR}{0\% (0/120; 95\% CI 0--3.1\%)}
\DeclareResult{R-FULL-RECALL}{100\% (1,440/1,440; 95\% CI 99.7--100\%)}
\DeclareResult{R-GEN-CAP-PROXY}{only 1/540 output equalled the frozen 1,600-token cap and 539/540 fell below it; no native finish reason was retained}
\DeclareResult{R-GEN-FENCED-FAILURE-TAXONOMY}{51 package-parse, 280 Python source-syntax, 99 other runtime, and 110 visible-contract failures}
\DeclareResult{R-GEN-FENCED-PARSE-YIELD}{90.6\% (489/540; 95\% CI 87.8--92.7\%)}
\DeclareResult{R-GEN-FUNCTIONAL-YIELD}{0\% (0/540; 95\% CI 0--0.7\%)}
\DeclareResult{R-GEN-HELDOUT-SUMMARY}{0/90 candidates parsed under the primary rule; normalization recovered 84/90, but 0/90 passed P0}
\DeclareResult{R-GEN-MODEL-FIT}{the prespecified mixed-effects conformance model was not estimable because the outcome had no variation}
\DeclareResult{R-GEN-PARSE-YIELD}{0\% (0/540; 95\% CI 0--0.7\%)}
\DeclareResult{R-GEN-RAW-N}{540/540}
\DeclareResult{R-HELDOUT-CLEAN-FPR}{0\% (0/20; 95\% CI 0--16.1\%)}
\DeclareResult{R-HELDOUT-DIFF}{74.6 percentage points}
\DeclareResult{R-HELDOUT-DIFF-CI}{72.5--76.7 percentage points}
\DeclareResult{R-HELDOUT-OP-DIFF}{75.0 percentage points}
\DeclareResult{R-HELDOUT-OP-DIFF-CI}{75.0--75.0 percentage points}
\DeclareResult{R-HELDOUT-OP-N}{160}
\DeclareResult{R-HELDOUT-RECALL}{100\% (240/240; 95\% CI 98.4--100\%)}
\DeclareResult{R-LATENCY-CONSTRUCTED}{4.37 s (95th percentile 4.74 s)}
\DeclareResult{R-LATENCY-GENERATED}{a median 1.78 s (95th percentile 1.80 s) across 489 rows}
\DeclareResult{R-N-CLEAN-ACCEPTED}{120}
\DeclareResult{R-N-CLEAN-CANDIDATES}{120}
\DeclareResult{R-N-GENERATED}{540}
\DeclareResult{R-N-INVALID-MUTANTS}{0}
\DeclareResult{R-N-SPECS}{30}
\DeclareResult{R-N-VALID-MUTANTS}{1,440}
\DeclareResult{R-P3-UNIQUE-FAMILIES}{six}
\DeclareResult{R-P3-UNIQUE-N}{680}
\DeclareResult{R-P4-RECORD-N}{395}
\DeclareResult{R-P4-RUNTIME-N}{zero}
\DeclareResult{R-P4-UNIQUE-N}{395}
\DeclareResult{R-P6-BASELINE}{identical on all 350 held-out rows to}
\DeclareResult{R-P6-HELDOUT-CLEAN-QUARANTINE}{0\% (0/20; 95\% CI 0--16.1\%)}
\DeclareResult{R-P6-HELDOUT-CLEAN-REJECT}{0\% (0/20; 95\% CI 0--16.1\%)}
\DeclareResult{R-P6-HELDOUT-DETECTION}{100\% (330/330; 95\% CI 98.8--100\%)}
\DeclareResult{R-P6-HELDOUT-QUARANTINE}{0\% (0/350; 95\% CI 0--1.1\%)}
\DeclareResult{R-P6-HELDOUT-RESIDUAL-RISK}{0\% (0/1,079 severity units)}
\DeclareResult{R-P6-SENSITIVITY}{selected the same thresholds and held-out decisions under both alternate severity-weight schemes and both quarantine-cost coefficients}
\DeclareResult{R-P6-THRESHOLDS}{quarantine at score 1 and reject at score 2}
\DeclareResult{R-PAIRED-TEST}{exact McNemar $p<0.001$; 1,075 discordant pairs, all favoring P0--P4}
\DeclareResult{R-PILOT-ACTIVATED}{400/400 (100\%; 95\% CI 99.0--100\%)}
\DeclareResult{R-PILOT-ATTEMPTS}{400}
\DeclareResult{R-PILOT-ISOLATED}{400/400 (100\%; 95\% CI 99.0--100\%)}
\DeclareResult{R-PILOT-REVERSED}{400/400 (100\%; 95\% CI 99.0--100\%) retained-parent reference records}
\DeclareResult{R-PRACTICE-SUPPORT-SUMMARY}{the controlled and bounded-repair evidence supports executable contracts and reason-preserving admission; P6 did not outperform simple fail-closed rejection, and queue prioritization remains scenario dependent}
\DeclareResult{R-PROMPT-PARSE-DIFF}{$-$7.0 percentage points}
\DeclareResult{R-PROMPT-PARSE-DIFF-CI}{$-$11.1 to $-$3.3 percentage points}
\DeclareResult{R-QUEUE-CLEAN-DIFF}{+0.04 time units (paired 95\% interval $-$0.36 to 0.42)}
\DeclareResult{R-QUEUE-DEFECT-DIFF}{$-$1.64 time units (paired 95\% simulation interval $-$4.32 to $-$0.52)}
\DeclareResult{R-QUEUE-P95}{8.43 time units (3.86--19.49)}
\DeclareResult{R-QUEUE-PARETO}{risk, risk-per-minute, and shortest-job-first as nondominated; FIFO is dominated}
\DeclareResult{R-QUEUE-RISK}{2.40 severity units (95\% simulation interval 0--10)}
\DeclareResult{R-QUEUE-SCENARIOS}{risk-per-minute reduced mean defective-package wait relative to FIFO in all seven runs, whereas the clean-package wait contrast changed sign}
\DeclareResult{R-QUEUE-THROUGHPUT}{0.796 packages per time unit (0.709--0.888)}
\DeclareResult{R-RECALL-DIFF}{74.7}
\DeclareResult{R-RECALL-DIFF-CI}{73.4--75.8}
\DeclareResult{R-REPAIR-CHANGED}{201/540 candidates}
\DeclareResult{R-REPAIR-DEDUP}{414 candidates, of which 23 passed P0 across 15 specifications}
\DeclareResult{R-REPAIR-DEFECTIVE}{112/112 P0-passing candidates}
\DeclareResult{R-REPAIR-DIFF}{86.6-percentage-point (specification-clustered 95\% CI 78.3--95.1)}
\DeclareResult{R-REPAIR-ELIGIBLE}{209/540 normalized candidates with syntactically valid Python source}
\DeclareResult{R-REPAIR-FAMILIES}{privacy 112, lifecycle 85, provenance 46, authority 4, and reliability 3}
\DeclareResult{R-REPAIR-HELDOUT}{17/90 candidates to pass P0; P0--P2 detected 0/17 and P0--P4 detected 17/17}
\DeclareResult{R-REPAIR-MULTIFAULT}{107/112 with multiple defect families}
\DeclareResult{R-REPAIR-P0}{20.7\% (112/540; 95\% CI 17.5--24.4\%)}
\DeclareResult{R-REPAIR-P2}{13.4\% (15/112; 95\% CI 8.3--20.9\%)}
\DeclareResult{R-REPAIR-P4}{100\% (112/112; 95\% CI 96.7--100\%)}
\DeclareResult{R-REPAIR-P4-MODES}{record-mediated evidence in 112/112 candidates and runtime-relation evidence in 35/112, with both modes in those 35}
\DeclareResult{R-ROBUSTNESS-RANGE}{a 74.2--75.3 percentage-point advantage across leave-one-archetype-out analyses}
\DeclareResult{R-RQ2-INTERPRETATION}{show that the contract and dynamic layers contribute complementary detections beyond conventional checks in the controlled stratum}
\DeclareResult{R-RQ3-INTERPRETATION}{supports transfer across the designed constructed holdout and archetype folds; the bounded-repair held-out result corroborates stage behavior but cannot establish unassisted generated-code transfer}
\DeclareResult{R-RUNTIME-PROBE}{240/240 runtime-only P4 rejections (120 per relation) after 0/240 P0--P2 rejections}
\DeclareResult{R-SHAPLEY-SUMMARY}{P1 12.1, P2 13.3, P3 47.2, and P4 27.4 percentage points (P0 0.0)}
\DeclareResult{T-ACC-N}{170}
\DeclareResult{T-ACC-P2}{0\%}
\DeclareResult{T-ACC-P4}{100\%}
\DeclareResult{T-AUT-N}{181}
\DeclareResult{T-AUT-P2}{0\%}
\DeclareResult{T-AUT-P4}{100\%}
\DeclareResult{T-DEP-N}{191}
\DeclareResult{T-DEP-P2}{100\%}
\DeclareResult{T-DEP-P4}{100\%}
\DeclareResult{T-FULL-CLEAN-CI}{100\% (96.9--100\%)}
\DeclareResult{T-FULL-CLEAN-N}{120}
\DeclareResult{T-FULL-CLEAN-PASS}{120}
\DeclareResult{T-FULL-MUT-CI}{100\% (99.7--100\%)}
\DeclareResult{T-FULL-MUT-N}{1,440}
\DeclareResult{T-FULL-MUT-VALID}{1,440}
\DeclareResult{T-LIF-N}{180}
\DeclareResult{T-LIF-P2}{0\%}
\DeclareResult{T-LIF-P4}{100\%}
\DeclareResult{T-PILOT-CLEAN-CI}{100\% (83.9--100\%)}
\DeclareResult{T-PILOT-CLEAN-N}{20}
\DeclareResult{T-PILOT-CLEAN-PASS}{20}
\DeclareResult{T-PILOT-MUT-CI}{100\% (99.0--100\%)}
\DeclareResult{T-PILOT-MUT-N}{400}
\DeclareResult{T-PILOT-MUT-VALID}{400}
\DeclareResult{T-PRI-N}{174}
\DeclareResult{T-PRI-P2}{0\%}
\DeclareResult{T-PRI-P4}{100\%}
\DeclareResult{T-PRO-N}{188}
\DeclareResult{T-PRO-P2}{0\%}
\DeclareResult{T-PRO-P4}{100\%}
\DeclareResult{T-REL-N}{182}
\DeclareResult{T-REL-P2}{0\%}
\DeclareResult{T-REL-P4}{100\%}
\DeclareResult{T-SEC-N}{174}
\DeclareResult{T-SEC-P2}{100\%}
\DeclareResult{T-SEC-P4}{100\%}

\title{EduPluginBench: Executable Assurance for AI-Generated Educational Plugins}

\author{Nizam Kadir}
\orcid{0000-0002-6725-1133}
\affiliation{%
  \institution{Singapore University of Technology and Design}
  \department{Science, Mathematics and Technology}
  \city{Singapore}
  \country{Singapore}}
\email{nizam\_kadir@mymail.sutd.edu.sg}

\begin{abstract}
Code-generation models can produce executable components, but compilation and
functional tests do not establish whether a component respects least privilege,
telemetry consent, evidence provenance, privileged-write authority, lifecycle
constraints, or bounded failure. We introduce EduPluginBench, an executable
benchmark and staged admission method for generated plugins in governed software
ecosystems. Matched configurations combine build and functional tests, static
security analysis, dependency and SBOM policy, executable domain contracts, and
sandboxed metamorphic execution with behavioral-record checks. The evidence is
hierarchical. On \result{R-N-VALID-MUTANTS} activation-checked first-order
mutants across \result{R-N-SPECS} specifications, P0--P4 increased
release-blocking-defect recall by \result{R-RECALL-DIFF} percentage points
(specification-clustered 95\% CI \result{R-RECALL-DIFF-CI}) over P0--P2, with
no observed rejection among \result{R-N-CLEAN-CANDIDATES} author-designed clean
references (95\% Wilson upper bound 3.1\%). A separately frozen transfer study
then evaluated \result{R-CONFIRM-DESIGN}: \result{R-CONFIRM-PARSE} parsed, but
\result{R-CONFIRM-P0-FULL}; downstream assurance estimands were therefore
undefined. An independently labelled Moodle study retained
\result{R-EXTERNAL-DESIGN}; \result{R-EXTERNAL-RESULTS}. These negative transfer
results prevent controlled contract consistency from being read as independent
real-defect effectiveness. An earlier \result{R-N-GENERATED}-generation
diagnostic remains separate: post-hoc bounded repair yielded
\result{R-REPAIR-P0} P0 passes, all nonconforming, with recall increasing from
\result{R-REPAIR-P2} to \result{R-REPAIR-P4}. Operational queue findings are
scenario-only. The artifact retains protocols, public-source provenance, raw
generations, row-level decisions, audits, analysis code, and exact reproduction
instructions.
\end{abstract}



\begin{document}
\maketitle

\section{Introduction}

Code-generating models have moved from synthesizing isolated functions to
editing repositories and coordinating software-engineering tasks~\cite{Chen2021Codex,Jimenez2024SWEBench}.
This change is particularly
consequential in extensible platforms. A generated component may be syntactically
valid and functionally plausible while also acquiring permissions, importing
dependencies, emitting telemetry, changing persistent state, or invoking
privileged services. Once admitted to a plugin ecosystem, such a component
becomes a maintained software artifact rather than a transient model response.

Educational platforms make the assurance problem unusually clear. Interoperable
learning tools carry identities and roles, read course data, and may write
assessment results through standardized services~\cite{OneEdTechLTI13}. A plugin
that returns the expected feedback can nevertheless overreach its role, expose a
synthetic learner identifier through telemetry, omit the source behind an
explanation, bypass a review transition, or retry a grade write after a partial
failure. These are software defects with domain semantics. They are not captured
completely by compilation, unit tests, generic static analysis, or dependency
scanning.

Existing evaluations reveal complementary parts of the problem. Functional
code-generation benchmarks use executable tests to establish whether programs
solve stated tasks~\cite{Chen2021Codex,Jimenez2024SWEBench,Pan2026RALBench}.
Secure-code studies
show that generated code can contain vulnerabilities and that functional success
does not imply security~\cite{Pearce2022Copilot,Bhatt2023CyberSecEval,
Peng2025CWEval}. Mutation testing supplies controlled faults and matched program
pairs, but the validity and representativeness of operators require explicit
evidence~\cite{Offutt1996Sufficient,Laurent2017PIT}. Plugin and extension
research, software supply-chain standards, and education interoperability
specifications each define relevant controls, yet they do not provide a single
executable method for deciding whether an AI-generated, domain-constrained
plugin should enter a governed ecosystem.

EduPluginBench addresses this gap as a software-engineering benchmark and
assurance method. A plugin package is evaluated against both conventional
software checks and an executable domain contract. Its evidence hierarchy
separates verified clean references, activation-checked first-order mutants,
exploratory and confirmatory model-generated candidates, independently labelled
Moodle vulnerable/fixed revisions, and synthetic operational traces. All
matched assurance configurations receive the same packages within a stratum;
no lower tier is pooled upward to enlarge a controlled-mutant claim.

The central claim is therefore narrower than universal ``safe code'' and
stronger than a feature demonstration: under explicit release contracts, the
method measures which defect classes a matched assurance pipeline exposes,
what clean-package harm it causes, and how well that evidence transfers beyond
the controlled mutant design. Perfect detection of an activation-validated
mutant establishes consistency with the encoded contract, not independent
real-world defect discovery. The unmodified-generation and public-defect strata
test that boundary separately and, in this study, return negative transfer
results.

The study addresses four research questions:

\begin{description}
  \item[RQ1---Detection.] What recall do individual and combined assurance
  stages achieve for release-blocking defects across security, dependency,
  privacy, authority, provenance, lifecycle, reliability, and accessibility,
  and what rejection do they impose on clean references?

  \item[RQ2---Complementarity.] Which defect families evade conventional
  software checks, and which assurance stages make unique detections after
  preceding stages are accounted for?

  \item[RQ3---Generalization.] How well do measured effects transfer to unseen
  specifications, mutation operators, plugin archetypes, unmodified
  model-generated packages, and independently labelled Moodle defects?

  \item[RQ4---Operational governance.] How does the calibrated admission policy
  trade residual risk against clean-package harm, and how do review priorities
  trade waiting time, backlog, and throughput in one predeclared base queue
  scenario and seven post-hoc sensitivity runs: one base-setting rerun and six
  one-factor variants?
\end{description}

\subsection{Contributions}

This article makes four ordered contributions. Each lower tier tests transfer
without enlarging the claim supported by the tier above it.

\begin{enumerate}
  \item \textbf{Controlled contract consistency.} We define an eight-family
  threat model, executable contract schema, and staged P0--P4 pipeline, then
  evaluate the matched contrast on clean references and activation-checked,
  isolated first-order mutants with specification-clustered uncertainty and
  held-out challenges.

  \item \textbf{Transfer to unmodified model-generated packages.} We preserve
  the earlier 540-candidate diagnostic and add a separately frozen 600-candidate
  study using current downloadable coding models, a larger output budget, an
  exact five-block interface, and no repair, resubmission, or content
  normalization. The observed zero P0 yield narrows rather than rescues the
  generated-code claim.

  \item \textbf{Independent external evidence.} We freeze public Moodle
  vulnerable/fixed revision pairs before applying a generic PHP Semgrep
  ruleset. Its zero-detection result exposes the transfer limit between
  benchmark-specific executable contracts and independently labelled,
  application-specific defects.

  \item \textbf{Operational and reproducibility support.} We evaluate admission
  and review-priority choices only as declared scenarios and supply a
  checksummed compendium in which numerical claims resolve to retained rows,
  analysis keys, and immutable configuration digests.
\end{enumerate}

Table~\ref{tab:evidence-map} makes the estimand boundaries explicit.

\begin{table}[t]
\caption{Evidence hierarchy and interpretation boundary.}
\label{tab:evidence-map}
\begin{tabular}{P{0.18\linewidth}P{0.25\linewidth}P{0.27\linewidth}P{0.20\linewidth}}
\toprule
Evidence tier & Unit & Supported claim & Does not establish \\
\midrule
Constructed clean/mutant & package or activation-validated mutant & consistency and clean-reference harm under the encoded contract & real-defect prevalence or independent discovery \\
Earlier generated diagnostic & raw or post-hoc repaired candidate & failure modes and selected-stratum downstream exposure & unassisted model reliability \\
Frozen confirmatory generation & unmodified model output & parse, P0, and full-conformance transfer yield & downstream recall when no row passes P0 \\
External Moodle pairs & public vulnerable/fixed announcement pair & transfer of the frozen generic PHP detector & validity of the six non-security contract families \\
Queue simulation & paired synthetic trace & trade-offs in declared scenarios & reviewer or institutional performance \\
\bottomrule
\end{tabular}
\end{table}

\section{Background and Related Work}

\subsection{Evaluation of AI-generated code}

HumanEval established execution-based functional evaluation for code language
models and demonstrated the importance of repeated sampling~\cite{Chen2021Codex}.
Repository-level benchmarks subsequently moved the unit of analysis from a
standalone function to changes that must coordinate across files, dependencies,
and tests~\cite{Jimenez2024SWEBench}. RAL-Bench evaluates runnable application
repositories with functional and non-functional tests, while CodeIF-Bench
evaluates compliance with verifiable coding instructions across turns
~\cite{Pan2026RALBench,Wang2025CodeIFBench}. Recent TOSEM studies
have examined natural-language code generation in development environments~\cite{Xu2022IDE},
multi-agent or self-collaborative generation~\cite{Dong2024SelfCollaboration},
output non-determinism~\cite{Ouyang2024Nondeterminism}, and the characterization and mitigation of
generated-code quality problems~\cite{Liu2024Refining}.

These benchmarks primarily ask whether generated code implements a requested
behavior or improves a software-quality measure. EduPluginBench asks a
different, release-oriented question: whether an executable package is
admissible under both software and domain constraints. The distinction matters
because a package may pass visible functional tests and still violate a
permission, provenance, telemetry, or lifecycle invariant. Generated candidates
therefore complement, rather than replace, controlled mutants in this study.

\subsection{Secure code generation and vulnerability evaluation}

Security studies have found vulnerable patterns in code produced by coding
assistants~\cite{Pearce2022Copilot} and have developed suites for insecure code,
attack assistance, and broader model risk~\cite{Bhatt2023CyberSecEval}. CWEval
argues for outcome-driven oracles that assess functionality and security
together~\cite{Peng2025CWEval}. SecRepoBench and RealSec-Bench move secure-code
evaluation into real repositories, and HardSecBench couples functional and
security checks for hardware and firmware tasks
~\cite{Shen2025SecRepoBench,Wang2026RealSecBench,Chen2026HardSecBench}.
User-centered evidence additionally shows that
access to an AI assistant does not itself guarantee secure solutions~\cite{Perry2023Insecure}.

EduPluginBench adopts executable outcomes and conventional SAST baselines, but
extends the evaluated property space. Security is one of eight defect families.
The remaining families include dependency integrity, data minimization,
role-constrained authority, evidence provenance, release lifecycle, bounded
failure, and accessible interaction. This cross-layer scope tests whether a
generic security pipeline is sufficient for governed plugin admission.

\subsection{Mutation testing and benchmark validity}

Mutation testing creates controlled program variants to evaluate the ability of
tests or analyses to expose faults. Classic TOSEM work studied sufficient mutant
operators and the coupling effect~\cite{Offutt1996Sufficient,Offutt1992Coupling}.
Subsequent work showed that practical mutation systems themselves require
careful assessment~\cite{Laurent2017PIT}. Real-fault benchmarks such as Defects4J
provide a complementary response to concerns that synthetic mutants may not
represent naturally occurring defects~\cite{Just2014Defects4J}.

The present benchmark treats a mutation operator as a measured instrument. An
operator--package application enters the effectiveness denominator only after a
label-separated construction oracle confirms the intended fault, the retained
immutable parent remains available with its original digest, and unrelated
family oracles retain their pre-mutation state. This retained-parent reference
and digest bookkeeping is a lineage safeguard, not independent mutation-validity
evidence and not inversion of the transformation.
Invalid and inapplicable applications remain in an audit table. First-order
mutants support controlled stage attribution; generated candidates and held-out
operators test whether conclusions are an artifact of that controlled design.

\subsection{Executable policy, supply-chain assurance, and plugin ecosystems}

Software assurance increasingly combines source analysis with provenance,
dependency, and build controls. The NIST Secure Software Development Framework
organizes practices across development and release~\cite{NIST2022SSDF}; SLSA
defines progressively stronger supply-chain
provenance~\cite{SLSA2026}; and CycloneDX represents software bills of
materials~\cite{CycloneDX16}. Domain constraints nevertheless remain application-specific.
Model-driven security demonstrates how policy can be represented and enforced
through software models~\cite{Basin2006ModelSecurity}, while runtime
verification connects temporal properties to monitors~\cite{Bauer2011RuntimeVerification}.

Extensible learning platforms already expose security-relevant services. LTI
1.3 uses OpenID Connect, signed JSON Web Tokens, and OAuth 2.0, while LTI
Advantage services govern role data, deep linking, and grade exchange~\cite{OneEdTechLTI13,OneEdTechSecurity}.
These standards define interoperability
and protocol conformance; they do not evaluate arbitrary generated plugin
implementations. LearnAdapt demonstrates a no-code authoring and governed
release workflow with syntax, API-boundary, SAST, review, deployment, and
telemetry controls~\cite{Kadir2026LearnAdapt}, but does not establish the
comparative detector effectiveness, benchmark validity, or held-out evidence
studied here. LearnAdapt motivates only the problem domain and high-level
governed-release setting. No code, schema, specification, dataset row, result,
table, figure, or evaluation record from that work is reused in EduPluginBench.
EduPluginBench uses education-platform role and service
boundaries as domain evidence while avoiding a claim of formal LTI
certification.

\subsection{Accessibility as a software quality constraint}

Accessibility defects are release-relevant software defects, not presentation
polish. WCAG 2.2 specifies perceivable, operable, understandable, and robust
interaction requirements~\cite{WCAG22}. Empirical TOSEM work reports that
accessibility remains unevenly integrated into software practice~\cite{Bi2022Accessibility}.
EduPluginBench includes mechanically testable
interface contracts---accessible names, keyboard operation, and non-color
status cues---while explicitly excluding claims about human usability.

\subsection{Contribution boundary}

The unresolved gap is not the absence of any one ingredient. Among the surveyed
work, we found no matched release-decision evaluation that connects conventional
software checks, executable domain obligations, controlled faults,
model-generated packages, clean-package harm, and operational consequences. Table~\ref{tab:boundary}
states that boundary without implying that every work in a category lacks
features outside its dominant evaluation design.

\begin{table}[t]
\caption{Feature-level comparison with the closest executable code-generation
benchmarks and the prior governed-authoring system. A filled circle denotes an
explicit evaluation dimension; a dash denotes that the dimension is outside the
stated evaluation design. ``Held-out'' denotes specification-, operator-, or
repository-level transfer evidence rather than an ordinary test split.}
\label{tab:boundary}
\begin{tabular}{P{0.18\linewidth}P{0.22\linewidth}cccccc}
\toprule
Work & Unit & Func. & Sec. & Domain & Mutation & Held-out & Admission \\
\midrule
RAL-Bench~\cite{Pan2026RALBench} & Application repository & $\bullet$ & -- & -- & -- & -- & -- \\
CWEval~\cite{Peng2025CWEval} & Code task & $\bullet$ & $\bullet$ & -- & -- & -- & -- \\
SecRepoBench~\cite{Shen2025SecRepoBench} & Repository completion & $\bullet$ & $\bullet$ & -- & -- & $\bullet$ & -- \\
RealSec-Bench~\cite{Wang2026RealSecBench} & Repository issue & $\bullet$ & $\bullet$ & -- & -- & $\bullet$ & -- \\
LearnAdapt~\cite{Kadir2026LearnAdapt} & Governed authoring workflow & $\bullet$ & $\bullet$ & $\bullet$ & -- & -- & $\bullet$ \\
EduPluginBench & Plugin package & $\bullet$ & $\bullet$ & $\bullet$ & $\bullet$ & $\bullet$ & $\bullet$ \\
\bottomrule
\end{tabular}
\end{table}

\section{System Model and Assurance Properties}

\subsection{Plugin package}

Let a plugin package be
\[
  p = \langle s, c, m, d, b, u, \ell \rangle ,
\]
where $s$ is a versioned specification, $c$ is source code, $m$ is a manifest,
$d$ is a locked dependency set, $b$ is executable behavior, $u$ is an
interaction description, and $\ell$ is lineage metadata. A specification
defines required functions and observable outputs together with permitted
roles, permissions, telemetry fields, evidence attributes, lifecycle
transitions, retry and timeout bounds, and interface obligations.

An assurance stage $A_j$ maps a package and specification to a decision,
diagnostics, and cost:
\[
 A_j(p,s) = \langle y_j, r_j, t_j \rangle ,
\]
where $y_j \in \{\mathrm{accept},\mathrm{reject},\mathrm{abstain}\}$, $r_j$ is
a set of machine-readable reasons, and $t_j$ is measured latency. The staged
pipeline is fail-closed when any mandatory stage rejects or abstains without an
approved review path.

Let $\mathcal{F}$ denote the eight defect families and let
$O_f(p,s)\in\{0,1\}$ be the hidden construction oracle for family $f$. A parsed
package is fully conforming when it passes P0 and no family oracle activates:
\[
 z(p,s)=I(A_0(p,s)=\mathrm{accept})
        \prod_{f\in\mathcal{F}}\bigl(1-O_f(p,s)\bigr).
\]
For detector stage $j$, write $d_j(p,s)=1$ when that stage rejects and define
the cumulative decision through stage $k$ as
$D_k(p,s)=\max_{0\leq j\leq k} d_j(p,s)$. The primary controlled-stratum
estimand is
\[
 \Delta_R=
 \Pr(D_4=1\mid m=1)-\Pr(D_2=1\mid m=1),
\]
where $m=1$ denotes an activation-validated first-order mutant. Generated
packages use $z$ and the family-oracle vector as separate outcomes; they are not
pooled with mutants to estimate $\Delta_R$.

\subsection{Threat model}

The generated package, its declared dependencies, and any requested network or
filesystem access are untrusted. The specification compiler, clean reference
implementations, hidden conformance tests, mutation registry, detector
configuration, sandbox launcher, and checksummed result store form the
trusted computing base. The adversary is not assumed to be a malicious model;
defects may arise from underspecification, insecure training examples, API
hallucination, stochastic decoding, or incomplete integration logic.

The primary adverse event is admission of a package containing at least one
release-blocking defect. A secondary adverse event is rejection of a clean
package, which reduces ecosystem yield and increases review load. Out of scope
are model-host compromise, hardware side channels, malicious maintainers after
admission, learning-outcome effectiveness, and usability.

\subsection{Contract families}

Table~\ref{tab:families} defines the eight predeclared defect families. Each
family has a construction oracle used for benchmark validity and a detector
implementation used by the evaluated pipeline. They are label-separated: the
detector implementations do not inspect mutation identifiers, expected stages,
or ground-truth labels, and regression tests confirm that their decisions are
invariant when those fields are stripped. The package object can nevertheless
carry lineage metadata, so this is an implementation and decision-invariance
claim, not interface-level non-receipt. The oracles and detectors are not
construct-independent because both operationalize the same specification and can
inspect the same public manifest or behavioral field.

\begin{table}[t]
\caption{Release-blocking defect families and executable evidence.}
\label{tab:families}
\begin{tabular}{P{0.16\linewidth}P{0.33\linewidth}P{0.41\linewidth}}
\toprule
Family & Example violation & Executable evidence \\
\midrule
Security & request-controlled evaluation or path escape &
Bandit, Semgrep, and frozen source-pattern rules \\
Dependency & vulnerable or unpinned package &
lockfile, SBOM, advisory snapshot, and version policy \\
Privacy & telemetry without consent or undeclared field &
manifest/behavioral-record checks; synthetic consent probe (diagnostic) \\
Authority & role or permission exceeds the specification &
capability-matrix/behavioral-record checks; denied-role probe (diagnostic) \\
Provenance & missing, stale, or inconsistent evidence &
schema checks and source-to-output trace relation \\
Lifecycle & illegal release transition or absent audit record &
declared-transition, audit-record, and rollback invariants \\
Reliability & unbounded retry, missing timeout, or side effect &
resource limits and metamorphic replay \\
Accessibility & missing name, keyboard path, or non-color cue &
declared-interface invariant checks \\
\bottomrule
\end{tabular}
\end{table}

\section{EduPluginBench Construction}

\subsection{Specification corpus}

The full corpus contains 30 executable specifications, six from each of five
archetypes: assessment, adaptive hinting, evidence retrieval, classroom
analytics, and learning-resource recommendation. An archetype varies in
functional purpose and privileged operations; specifications within an
archetype vary in data shape, allowed roles, state transitions, evidence
requirements, and failure policy.

Each specification contains:
\begin{enumerate}
  \item a natural-language task statement and typed plugin API;
  \item a JSON Schema for manifest and output records;
  \item an explicit capability matrix for roles, data, and write operations;
  \item a telemetry allowlist and consent preconditions;
  \item provenance fields and freshness rules;
  \item a lifecycle transition system;
  \item deterministic functional examples visible to generation models;
  \item hidden conformance and adversarial tests; and
  \item mechanically checkable accessibility and resource bounds.
\end{enumerate}

Specifications use synthetic identities, submissions, events, course content,
and evidence. No production records or participant data enter the corpus.
Identifiers and text are generated from frozen templates, then committed before
model inference.

\subsection{Clean reference packages}

Four structurally distinct source-implementation families instantiate each
specification, for 120 clean candidates. The families differ in control flow,
helper decomposition, and data construction while exposing the same API\@. A package is
accepted as clean only if it passes visible functional tests, the hidden
conformance suite, all family oracles, deterministic replay, dependency policy,
and a duplicate-digest check. Failure excludes the package and all descendants;
the exclusion is recorded.

Clean packages serve two purposes. They estimate stage-specific false-positive
rates, and they provide immutable parents for controlled mutants. They do not
estimate the prevalence of clean AI-generated software.

\subsection{Mutation registry}

The full registry targets 40 operators, five per family. Operators are derived
from the specification vocabulary and software mechanisms, not detector error
inspection. Each record contains a stable identifier, precondition, exact
transformation, severity, family, activation oracle, retained-parent reference,
expected observability boundary, and rationale.

For a parent $p$ and operator $\mu$, a mutant $p'=\mu(p)$ is valid only when:
\begin{align}
  H_f(p,s)&=0, & H_f(p',s)&=1, \\
  H_g(p',s)&=H_g(p,s) && \forall g\neq f,
\end{align}
where $H_f$ is the construction oracle for the intended family. All 40
operators are defined as total transformations over the canonical package
schema. A post-hoc construction audit therefore exercises all 4,800
operator--parent pairs for applicability, activation, isolation, and preservation
of the immutable parent reference and digest. This bookkeeping confirms that
mutants remain linked to unchanged retained parents; it is not counted as
independent validity evidence and does not claim an inverse for $\mu$. Before detector
execution, a frozen severity-balanced design retains 12 pairs per clean package
for the primary effectiveness study; remaining valid pairs are recorded as
unsampled rather than invalid. One operator per family is reserved from
development and calibration for the predeclared held-out-operator analysis.

\subsection{Model-generated candidates}
\label{sec:generated-candidates}

The generated stratum crosses 30 specifications, three downloadable model
families, two prompt conditions, and three random seeds, producing a target of
540 raw candidates. The selected models are
Qwen2.5-Coder-7B-Instruct, DeepSeek-Coder-6.7B-Instruct, and
Phi-4-mini-instruct~\cite{Hui2024QwenCoder,Guo2024DeepSeekCoder,Abouelenin2025Phi4Mini}.
Their exact repository commit hashes and licenses are frozen
in the model manifest; each commit jointly pins the model, tokenizer, and chat
template files loaded by the notebook. Four-bit NormalFloat (NF4) inference
permits execution on a standard Colab GPU; the assigned accelerator and package
versions are
recorded. The model-specific compatibility pins use Transformers 4.57.1 for
Qwen and DeepSeek and 4.53.3 for Phi-4; the remaining inference-package
versions are shared and retained in every raw row and run manifest.

The \emph{interface-contract-only} condition includes the task, package schema,
API, and visible examples. The \emph{assurance-aware-contract} condition adds
the same executable obligations later enforced by the domain contract, without
exposing hidden tests or mutation operators. Generation uses temperature 0.2,
top-$p$ 0.95, a 1,600-token limit, and seeds 1701--1703. Raw text and token IDs
are stored before extraction. No repair is applied to the primary candidate;
parse failures and refusals remain outcomes.

The primary extractor accepts exactly one bare JSON object, as requested in the
prompt, and treats Markdown fences or surrounding prose as interface-contract
failures. After the primary evaluation produced no parseable package, we added
a transparently post hoc format-normalization sensitivity. It extracts the
payload of exactly one code fence labelled \texttt{json}, without changing any
byte inside that payload or synthesizing, deleting, or coercing fields. Zero or
multiple matching fences, invalid JSON, and missing or mistyped fields still
fail. The strict result remains primary, and the amendment record timestamps
the sensitivity after observation of the universal strict-format failure.

A second post-hoc analysis separates generation-protocol failure from assurance
failure. First, retained inference records are audited for completion status,
input/output token counts, and finish reasons. Because the records do not expose
a tokenizer-native finish reason, equality with the 1,600-token cap is treated
only as a conservative truncation proxy. Second, a frozen bounded source-repair
policy operates only on syntactically valid source recovered by the fenced-JSON
normalizer. The policy was designed after aggregate failure classes and recurrent
candidate source forms were inspected; it therefore remains outcome-informed
and post hoc. It permits three transformations anchored in the public schema and
visible examples:
\texttt{"success"} to \texttt{"ok"} for the returned status literal, mapping
access for three provenance fields incorrectly written as context attributes,
and \texttt{strip()} on a directly returned event value. It cannot synthesize
source, alter manifests, dependencies, behavior or interface metadata, suppress
exceptions, or use hidden-test feedback. We therefore describe it as
public-contract-only and hidden-test-independent, not as independent of the
observed candidate corpus. Whenever an adapter fires, Python's
\texttt{ast.unparse} serializes the transformed syntax tree; this normalizes
formatting and removes comments in addition to the enumerated semantic edit.
The unchanged strict endpoint remains the generation
result; bounded repair supplies a clearly labelled sensitivity stratum for
testing downstream assurance stages.

Generated packages may contain zero, one, or multiple defect families. Their
family labels derive from reference-oracle functions that are label-separated
from the detector stages but not construct-independent. For properties
represented in manifest or behavioral records,
the reference oracle and detector necessarily inspect some of the same public
fields; Section~\ref{sec:threats} therefore treats generated-stratum agreement as corroboration,
not as construct-independent ground truth. Ambiguous cases are quarantined under a
predeclared rule and reported separately rather than assigned by detector
output. This stratum estimates functional yield and observed defect composition
under the sampled generation regimes; it is not pooled with single-fault
mutants for the primary recall estimate.

\subsection{Confirmatory unmodified generation transfer}
\label{sec:confirmatory-generation}

The targeted confirmatory study was frozen under identifier
\texttt{EDUPLUGINBENCH-CONFIRM-001} before either model output existed. Its
complete $30\times2\times2\times5$ design crosses every specification with
Qwen3-Coder-30B-A3B-Instruct and Devstral-Small-2507
\cite{Qwen3CoderModel,DevstralModel}, the
\texttt{scaffold\_contract\_only} and
\texttt{scaffold\_assurance\_aware} prompts, and seeds 2711--2715. Exact model
revisions are retained. Generation uses temperature 0.2, top-$p$ 0.95, a
4,096-new-token ceiling, and NF4 inference on the recorded Colab accelerator.
This is a transfer study, not a model ranking.

Both prompts require the same ordered five-block scaffold: manifest,
dependencies, behavior, interface, and Python source. The first four payloads
must be JSON objects; source bytes are preserved. The frozen parser permits raw
scaffold text or one outer fence labelled \texttt{text} or
\texttt{plaintext}; it rejects commentary, unlabelled fences, additional bytes,
missing blocks, and mistyped payloads. It never synthesizes, deletes, coerces,
or edits package content. Failed candidates are neither repaired nor
resubmitted. Primary outcomes are parse yield, P0 yield, and full P0--P4
conformance yield. Downstream recall and clean-rejection are conditioned on an
unmodified P0 pass and are reported as undefined when that set is empty.

\subsection{Independent Moodle vulnerability/fix pairs}
\label{sec:external-moodle}

The external security-facing study begins from the 18 public Moodle core
security announcements MSA-26-0012 through MSA-26-0029
\cite{MoodleSecurityAnnouncements}. Published CVE and MDL identifiers supply
labels independently of EduPluginBench. For each announcement, the acquisition
script resolves the Moodle GitHub-mirror commit whose message contains the MDL
identifier, labels changed PHP files at its first parent as vulnerable, and
labels their counterparts at the fixing revision as fixed. Merge commits,
ambiguous commit searches, unavailable sources, and non-PHP-only changes are
excluded without replacement and retained with reason codes. The corpus
manifest, source bytes, paths, parent/fix revisions, and hashes were locked
before detector execution.

The frozen detector is Semgrep 1.172.0 with the downloaded generic community
\texttt{p/php} ruleset and metrics disabled \cite{SemgrepPHP}. A pair is detected
when at least one security finding occurs in its vulnerable revision;
fixed-revision rejection estimates matched false alarms. Secondary directional
reduction requires a vulnerable-revision finding to disappear in the fixed
revision. The historical PHP is scanned as inert text and never executed. This
study tests only transfer of the generic security-facing detector, not the six
non-security domain-contract families.

\subsection{Data partitioning and leakage control}

Five specifications---one per archetype---form the final held-out set. They are
selected by a frozen seed before candidate generation and do not inform detector
thresholds. The remaining 25 specifications comprise 15 development and 10
calibration specifications. Generated outputs, extracted packages, and
descendants inherit their specification partition.

Additional stress tests withhold one mutation operator per family, perform five
leave-one-archetype-out analyses, and stratify generated results by prompting
condition. Canonical package and source digests enumerate exact duplicates for
audit. Detectors receive package
content and the public specification. Although the package object can contain
lineage fields, detector implementations do not inspect the model identity,
prompt condition, mutation identifier, parent identifier, expected stage, split,
or ground-truth label; stripping the available lineage labels leaves decisions
unchanged in regression tests.

The predeclared eight-operator challenge applies those operators only to the five
held-out specifications and is therefore a joint operator-and-specification
challenge. A post-hoc factorial challenge applies the same eight previously
reserved operators to all 120 clean parents across all 30 specifications. This
second design removes the original operator--specification confound but remains
a fixed-set robustness result rather than a random sample of possible operators.

\section{Assurance Pipeline}

Table~\ref{tab:pipelines} defines cumulative configurations. P0--P2 represent
the conventional baseline; P3 and P4 add domain and behavioral assurance. P6
is the calibrated admission policy; the review-queue scenarios are evaluated
separately rather than being populated from P6 outcomes. P5 is omitted because
the study does not measure human review.

\begin{table}[t]
\caption{Matched cumulative assurance configurations.}
\label{tab:pipelines}
\begin{tabular}{p{0.08\linewidth}p{0.82\linewidth}}
\toprule
ID & Configuration \\
\midrule
P0 & Package schema, syntax/build, visible functional tests, and deterministic packaging \\
P1 & P0 plus pinned Bandit and Semgrep static security rules \\
P2 & P1 plus lockfile policy, CycloneDX SBOM, and a frozen vulnerability-advisory snapshot \\
P3 & P2 plus executable authority, telemetry, provenance, lifecycle, resource, and accessibility contracts \\
P4 & P3 plus network-isolated execution, normalized scenario traces, metamorphic relations, and behavioral-record checks \\
P6 & Calibrated fail-closed composition with an explicit abstain/quarantine path \\
\bottomrule
\end{tabular}
\end{table}

\subsection{P0: package and functional checks}

P0 validates package and manifest shape, parses and compiles source, checks the
dependency record, executes the visible example twice in a digest-pinned Python
3.12 container, and checks fixed-input determinism. The container has no
network, a read-only root, bounded temporary storage, CPU, memory, process, and
wall-clock limits, dropped capabilities, and an unprivileged user. Timeouts,
crashes, missing entrypoints, malformed output, and visible-example failures
cause rejection. P0 establishes basic functionality but does not inspect hidden
domain obligations. The 256 MiB memory cap is enforced, but peak resident memory
is not measured; timing is descriptive, and P0 carries the cost of the shared
sandbox invocation whose traces P4 later reuses.

\subsection{P1: static application security testing}

P1 executes pinned versions and rule sets of Bandit and Semgrep. Rules cover
request-controlled evaluation, command and path injection, unsafe
deserialization, hard-coded credentials, insecure temporary files, weak
cryptography, and unrestricted network clients. Severity mapping and suppression
policy are frozen before held-out execution. In this benchmark, every finding
from the frozen rules is release blocking. Unexpected tool return codes or
non-JSON output also reject fail closed; raw findings and return codes remain
available for audit.

\subsection{P2: dependency and supply-chain policy}

P2 requires exact dependency versions, generates a CycloneDX SBOM, and evaluates
the dependency record against an archived OSV snapshot and name/source policy.
Pinned \texttt{pip-audit} output is retained as a contemporaneous diagnostic but
does not determine the historical binary outcome. This separation prevents
later advisory changes from rewriting the recorded decision.

\subsection{P3: executable domain contracts}

P3 evaluates deterministic contract checks derived from the specification:
\begin{itemize}
  \item a capability check compares declared roles, permissions, and behavioral
  authority records with the role--permission matrix;
  \item a telemetry check evaluates consent and field minimization records;
  \item a provenance check relates declared output evidence to freshness metadata;
  \item a lifecycle check accepts only permitted declared transitions and
  requires an audit record;
  \item a resource contract checks retry and timeout declarations; and
  \item an interface contract checks accessible names, keyboard paths, and
  redundant status cues.
\end{itemize}
Contract evaluation is deterministic. The mutation registry, labels, and
expected stages are unavailable to the evaluator.

\subsection{P4: dynamic and metamorphic assurance}

P4 reuses the pre-execution gate and digest-pinned container from P0. The runner
executes normal, fixed-replay, irrelevant-field, consent-withdrawal,
stale-evidence, denied-role, and partial-failure probes and stores a normalized
record of returns and exceptions. Two relations have uniquely specified
outcomes and therefore determine rejection: an irrelevant event field must not
change the result, and the returned evidence version and retrieval time must
track the stale-evidence input. The remaining probes are preserved as
diagnostics because this API does not specify one universal response for them.
For normalized return trace $T$ and irrelevant field $q$, the implemented
relations are
\[
 T(e,c)=T(e\oplus q,c)
 \quad\text{and}\quad
 \operatorname{evidence}(T(e,c'))=
   \langle c'.\mathrm{version},c'.\mathrm{retrieved\_at}\rangle .
\]
Exceptions and malformed returns remain observable outcomes rather than being
coerced into equality.

P4 also checks behavioral records for runtime-facing constraints that cannot be
observed from the minimal \texttt{run}(\texttt{event}, \texttt{context}) return channel: telemetry
field minimization and endpoint, caller authority, emitted-evidence consistency,
network use, determinism, and idempotency. This is record-mediated assurance,
not complete system-call tracing; the distinction is explicit in the validity
analysis and artifact.

Thus P4 contains three evidence classes: the two rejection-bearing executable
relations above; consent, denied-role, and partial-failure probes retained only
as diagnostics; and static/declarative behavioral-record checks that can reject.
Results attribute record and executable-relation findings separately.

Table~\ref{tab:worked-trace} gives one retained decision trace. It is
illustrative, not an additional analysis: the activation-validated
\texttt{PRI-NO-CONSENT} mutant for
\texttt{assessment.formative\_quiz} passed the conventional stages, while P3
rejected its explicit telemetry-without-consent violation. P4 added no finding
for this package, and the cumulative admission decision remained reject.

\begin{table}[t]
\caption{Worked P0--P4 trace for one controlled privacy mutant.}
\label{tab:worked-trace}
\begin{tabular}{P{0.11\linewidth}P{0.14\linewidth}P{0.60\linewidth}}
\toprule
Stage & Decision & Retained reason \\
\midrule
P0 & accept & schema, build, visible example, and repeat execution pass \\
P1 & accept & no frozen SAST finding \\
P2 & accept & dependency and SBOM policy pass \\
P3 & reject & \texttt{privacy: telemetry enabled without consent} \\
P4 & no added finding & neither runtime relation nor behavioral-record rule adds a reason \\
\midrule
Cumulative & reject & first release-blocking finding occurs at P3 \\
\bottomrule
\end{tabular}
\end{table}

\subsection{P6: calibrated admission}

P6 maps stage outputs to admit, quarantine, or reject. Calibration minimizes
severity-weighted residual risk subject to two constraints fixed in the full
protocol: clean-package rejection at most 10\% and a quarantine budget of at
most 20\%. The frozen stage weights are 12, 5, 4, 3, and 2 for P0 through P4;
thus a nonfunctional P0 failure cannot be offset by later non-rejections.
Defect-family severity weights are 5 for security, privacy, and authority; 4
for dependency, provenance, and lifecycle; and 3 for reliability and
accessibility. These ordinal weights encode the study's release scenario, not a
universal risk scale. The selected policy is frozen before final held-out
evaluation. No threshold is retuned on the held-out specifications.

For $S(p)=\sum_{j=0}^{4}a_jd_j(p,s)$, calibration enumerates integer admit,
quarantine, and reject thresholds, retains only policies satisfying both harm
bounds, and minimizes admitted severity-weighted risk with deterministic
tie-breaking. A post-hoc comparison with reject-on-any-P0--P4-finding asks
whether calibration changes held-out decisions; it does not retune the policy.

\section{Empirical Study}

\subsection{Study overview}

Table~\ref{tab:studies} separates benchmark validation from effectiveness and
operational questions. This separation prevents perfect activation of a
constructed mutant from being presented as evidence of real-world detector
effectiveness.

\begin{table}[t]
\caption{Studies, units, and estimands.}
\label{tab:studies}
\begin{tabular}{P{0.09\linewidth}P{0.30\linewidth}P{0.25\linewidth}P{0.26\linewidth}}
\toprule
Study & Purpose & Unit & Principal estimand \\
\midrule
S0 & Pilot mechanics and go/stop gates & operator--parent pair & activation, retained-parent lineage, isolation \\
S1 & Reference benchmark validity & specification and clean package & clean acceptance and mutant validity \\
S2 & Matched detector effectiveness & valid first-order mutant & recall difference and stage complementarity \\
S3 & Exploratory generation diagnostics & raw and bounded-repair generated candidate & yield, selected-stratum violation composition, downstream recall \\
S4 & Generalization and ablation & held-out specification/operator/archetype and post-hoc runtime probe & transfer and component capability \\
S5 & Operational governance & paired simulated trace & residual risk, delay, yield, throughput \\
S6 & Confirmatory generation transfer & unmodified model-generated candidate & parse, P0, full-conformance, and conditional assurance yield \\
S7 & External security transfer & public Moodle vulnerable/fixed announcement pair & vulnerable recall, fixed rejection, and directional reduction \\
\bottomrule
\end{tabular}
\end{table}

\subsection{Design-size rationale}

Study size follows coverage constraints fixed before full execution. Thirty
specifications provide six distinct tasks in each of five archetypes and
permit one held-out specification per archetype; four implementation families
vary clean-package structure; and at most 12 severity-balanced valid mutants
per clean parent provide operator coverage without allowing prolific operators
to dominate. The 540 generated candidates are the complete
$30\times3\times2\times3$ specification--model--prompt--seed design, not a
post-result subsample. S6 adds the complete, independently frozen
$30\times2\times2\times5=600$ unmodified-candidate design. S7 seeds 18 public
announcements and retains every pair satisfying its predeclared source-history
eligibility rules without replacement. Because the constructed corpus is a designed benchmark
rather than an independent and identically distributed (iid) prevalence
sample, a conventional row-level power
calculation would misstate its information. Specification-clustered intervals
and the held-out analyses expose the achieved precision and transfer limits.

\subsection{Pilot and protocol lock}

Before full-scale construction, S0 uses 10 specifications, 20 clean packages,
20 mutation operators, P0--P4, one held-out specification, and 1,000 queue
replications per policy. The protocol is SHA-256 locked before the recorded run.
Scaling requires at least 90\% activation, a matching retained-parent digest for every activated
mutants, isolation of intended faults, detection gain in at least three
non-security families, at least a 10-percentage-point recall gain over P0--P2,
a clean false-positive rate below 10\%, and a viable held-out split. Pilot
observations validate mechanics only and are excluded from confirmatory
full-study estimates.

\subsection{Hypotheses}

The confirmatory hypotheses are:
\begin{description}
  \item[H1.] P0--P4 has higher release-blocking-defect recall than P0--P2 on
  valid first-order mutants.
  \item[H2.] P3 contributes non-zero unique detection in at least three of the
  six non-security, non-dependency domain families.
  \item[H3.] P6 satisfies the preregistered clean-package rejection bound on
  held-out specifications.
  \item[H4.] The recall advantage of P0--P4 over P0--P2 remains positive for
  held-out specifications and held-out operators.
  \item[H5.] Assurance-aware prompting changes full-conformance yield relative
  to interface-contract-only prompting.
\end{description}
RQ4 is evaluated as a multi-objective scenario analysis rather than a null
hypothesis test.

\subsection{Outcome measures}

For valid mutants, primary effectiveness is family-stratified recall. For clean
packages, the primary harm measure is false-positive rate. For generated
candidates, outcomes are parse yield, functional yield, clean yield,
multi-family defect count, and rejection or abstention by stage. Precision is
reported only within strata whose constructed class balance is explicit; it is
not interpreted as deployment prevalence.

Let $\mathcal{F}_i$ be the confirmed defect families for candidate $i$ and
$s_i=\sum_{f\in\mathcal{F}_i}w_f$ its predeclared severity weight. Residual
severity-weighted risk for an admission policy $\pi$ is
\[
  R(\pi)=
  \frac{\sum_{i} s_i I(g_i=1) I(\pi_i=\mathrm{admit})}
       {\sum_i s_i I(g_i=1)},
\]
where $g_i$ denotes any confirmed defect; candidates with multiple confirmed
families contribute the sum of their family weights. Operational outcomes
include clean yield, quarantine rate,
stage latency, simulated review waiting time, tail delay, throughput, and the
severity sum of defective jobs unfinished at the final-arrival horizon. This
last queue measure is distinct from P6's admitted-risk fraction.

\subsection{Statistical analysis}

Counts and proportions are reported with Wilson 95\% intervals. Primary
uncertainty uses a cluster bootstrap that resamples specifications and retains
all packages and descendants within a selected specification. Each clustered
contrast uses 10,000 percentile-bootstrap replications; contrast-specific
random seeds are frozen in the summary artifacts. The primary contrast is the
absolute recall difference between cumulative P0--P4 and P0--P2. Because
configurations are matched and nested, discordant counts and an exact McNemar
test are reported alongside the bootstrap interval. The McNemar calculation is
a row-level matched diagnostic; inferential interpretation rests on the
specification-clustered interval because descendants of one specification are
not independent.

For the generated stratum, a mixed-effects logistic regression models full
conformance with fixed effects for model family, prompt condition, and archetype,
plus a random intercept for specification. Odds ratios and intervals supplement
the clustered absolute prompt-condition contrast. The pinned Statsmodels 0.14.5
fit uses variational Bayes, a zero-centered Gaussian fixed-effect prior with
standard deviation 2, and a zero-centered Gaussian prior with standard deviation
1 on the log random-effect standard deviation; reported 95\% intervals are
posterior normal approximations. A singular or otherwise unusable
random-intercept fit triggers the predeclared
specification-clustered logistic GEE fallback, and the trigger is retained in
the summary artifact. If the binary outcome has no variation, the model is
reported as non-estimable rather than forcing a continuity correction.

S6 reports Wilson intervals for parse, P0, and full-conformance yields. Its
prompt contrasts first average within specification and then use 10,000 frozen
specification-cluster bootstrap resamples. Assurance recall and clean-rejection
are calculated only among unmodified P0-passing candidates; an empty eligible
set yields an undefined estimand, not an imputed zero. S7 uses the announcement
pair as the unit: vulnerable and fixed rejection receive Wilson intervals,
matched discordance receives the exact McNemar test, and directional reduction
is a prespecified secondary count. The post-hoc miss audit is descriptive and
does not alter any S7 decision.

Holm-adjusted probabilities are reported across the two predeclared secondary
row-level matched diagnostics. Effect sizes and specification-clustered
intervals, rather than statistical significance alone, determine
interpretation. Missing stage observations are never imputed:
infrastructure failures are rerun under the same run identifier when safe; a
repeated failure becomes an abstention and is included in the fail-closed
analysis. All exclusions are enumerated by reason.

\subsection{Robustness and ablation analyses}

Predeclared robustness analyses repeat the primary contrast on:
\begin{enumerate}
  \item five held-out specifications;
  \item eight held-out mutation operators;
  \item five leave-one-archetype-out folds;
  \item generated candidates by model, prompt condition, and archetype;
  \item the P0-passing generated subset for detector recall and clean harm; and
  \item generated results after seeded-representative deduplication by
  canonical package, exact source, and source-token 5-gram Jaccard thresholds
  0.90 and 0.95;
  \item fail-closed versus infrastructure-complete-case handling;
  \item alternate severity weights for the calibrated admission policy; and
  \item P6 quarantine-cost coefficients 0.25 and 1.00 under the unchanged
  calibration constraints.
\end{enumerate}

Post-hoc analyses are reported separately from this predeclared set. The
single-fenced-JSON decomposition assigns each raw candidate to exactly one of
package-parse failure, Python source-syntax failure, other runtime failure,
visible-contract failure, or P0 pass. The bounded-repair analysis then reports
P0 yield, family-oracle prevalence, cumulative P0--P2 and P0--P4 recall,
P4 record/runtime attribution, exact-source deduplication, and a
specification-clustered recall contrast. Generation diagnostics inspect status
and token counts without inferring an unavailable finish reason. Operator
sensitivities comprise the 4,800-pair construction audit, an eight-operator
factorial challenge across all 30 specifications, and two source-level runtime
capability probes applied once to each of 120 clean parents. None modifies the
primary parser, primary mutant sample, or confirmatory estimand.

Ablations remove P1, P2, P3, or P4 from the cumulative pipeline while holding
all other decisions constant. Shapley-style order averaging is reported as a
descriptive sensitivity because unique detection in a cumulative order can
depend on stage position.

\subsection{Operational Monte Carlo study}

S5 uses a discrete-event, single-server queue with paired traces: FIFO, risk
priority, risk-per-expected-minute, and shortest expected review receive the
same 300 synthetic jobs per replication. The base scenario fixes exponential
arrival rate 0.8, defect prevalence 0.25, class-conditional Beta risk signals,
lognormal service demand, and master seed 40,731. Ten thousand replications
report paired waiting-time, tail, backlog, unfinished-severity, and throughput
contrasts with a Pareto analysis. Seven post-hoc 2,000-replication runs vary one
arrival, prevalence, or service-dispersion parameter at a time. All times are
dimensionless scenario units; no trace is sampled from people, P6 rows, SUTD,
or another institution.

\subsection{Reproducibility controls}
\label{sec:reproducibility-controls}

Recorded runs include a study identifier, run identifier, protocol-lock digest,
UTC timestamp, platform and tool versions, accelerator metadata when applicable,
random seeds, and input-file digests. A final artifact manifest adds SHA-256
digests for every distributed file. Raw generation text, packages, detector
findings, normalized traces, timing rows, and exclusion rows are retained;
derived tables and figures are regenerated from row-level data. The strict
manuscript build fails if a result key is unresolved.

OpenAI Prism and Codex assisted with implementation scaffolding, analysis and
verification code, figure production, literature discovery, manuscript drafting,
and adversarial editorial review. Their outputs were not treated as evidence or
authorship: experimental claims were retained only when traceable to frozen raw
rows and executable analyses, proposed code changes were exercised by the test
and clean-room workflows, literature statements were checked against cited
sources, and numerical manuscript values were resolved through the evidence
registry. The downloadable models in
Section~\ref{sec:generated-candidates} served only as experimental code
generators and did not supply oracle labels or detector decisions. Nizam Kadir
reviewed every retained change, is the sole human author, directed the study,
and accepts responsibility for all research and writing decisions.

\section{Results}

\subsection{Benchmark validity (S0--S1)}

The pilot attempted \result{R-PILOT-ATTEMPTS} operator--parent applications.
\result{R-PILOT-ACTIVATED} activated the intended fault, and
\result{R-PILOT-ISOLATED} remained isolated to one family. The retained immutable
parent reference and matching digest were recorded for
\result{R-PILOT-REVERSED} activated mutants; this is lineage bookkeeping, not an
inverse-transformation result or independent validity evidence. In the full
benchmark, \result{R-N-CLEAN-ACCEPTED} of \result{R-N-CLEAN-CANDIDATES} reference
packages passed the complete clean oracle. The operator audit retained
\result{R-N-VALID-MUTANTS} valid first-order mutants and recorded
\result{R-N-INVALID-MUTANTS} invalid or inapplicable attempts
(Table~\ref{tab:validity-results}). The separate post-hoc all-pair audit
\result{R-ALL-PAIR-AUDIT}; thus the 1,440-mutant denominator is the frozen
balanced effectiveness sample, not the number of transformations attempted.

\begin{table}[t]
\caption{Benchmark validity counts and Wilson 95\% intervals.}
\label{tab:validity-results}
\begin{tabular}{lrrr}
\toprule
Stratum & Attempted & Accepted/valid & Rate (95\% CI) \\
\midrule
Pilot clean packages & \result{T-PILOT-CLEAN-N} & \result{T-PILOT-CLEAN-PASS} & \result{T-PILOT-CLEAN-CI} \\
Pilot mutations & \result{T-PILOT-MUT-N} & \result{T-PILOT-MUT-VALID} & \result{T-PILOT-MUT-CI} \\
Full clean packages & \result{T-FULL-CLEAN-N} & \result{T-FULL-CLEAN-PASS} & \result{T-FULL-CLEAN-CI} \\
Full mutations & \result{T-FULL-MUT-N} & \result{T-FULL-MUT-VALID} & \result{T-FULL-MUT-CI} \\
\bottomrule
\end{tabular}
\end{table}

\subsection{Matched assurance effectiveness (RQ1)}

\begin{figure}[t]
  \centering
  \includegraphics[width=\linewidth]{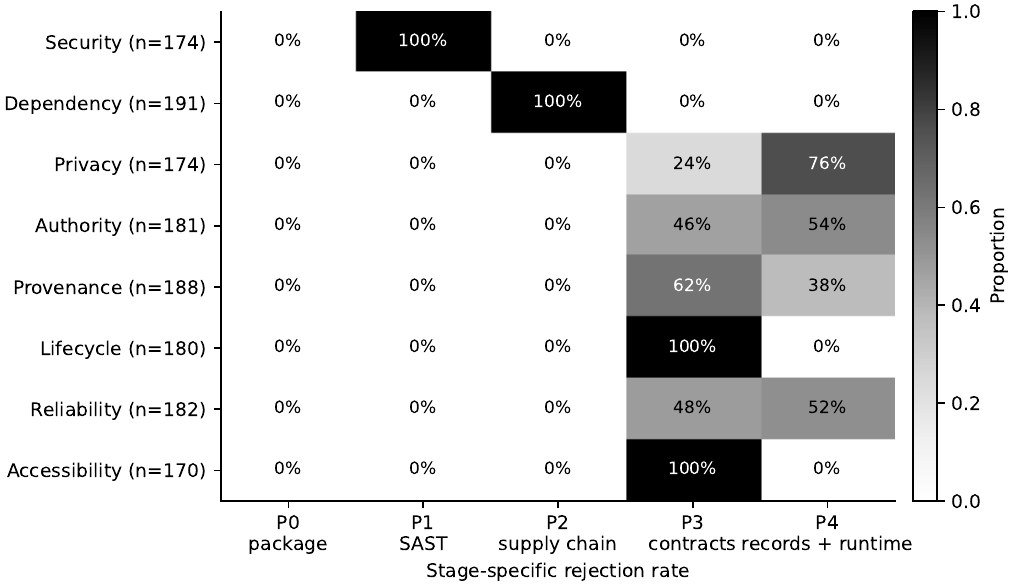}
  \caption{Stage-specific rejection rates on controlled mutants. Cells show
  the proportion rejected at that stage, not cumulative recall; row labels give
  the frozen family denominators. The matrix makes stage complementarity and
  any same-package overlap visible without duplicating
  Table~\ref{tab:family-results}.}
  \Description{Eight-by-five grayscale matrix with defect families as rows and
  stages P0 through P4 as columns. Security concentrates at P1, dependency at
  P2, and the six domain families at P3 and P4. All primary P4 detections are
  record-mediated and the runtime-specific contribution is zero; percentages
  appear in cells.}
  \label{fig:constructed-stage-matrix}
\end{figure}

On valid first-order mutants, cumulative P0--P2 detected
\result{R-CONVENTIONAL-RECALL}, compared with \result{R-FULL-RECALL} for
P0--P4, an absolute difference of \result{R-RECALL-DIFF} percentage points
(clustered 95\% CI \result{R-RECALL-DIFF-CI};
\result{R-PAIRED-TEST}). Clean-package cumulative false-positive rates were
\result{R-CONVENTIONAL-CLEAN-FPR} and \result{R-FULL-CLEAN-FPR},
respectively. These are 0 observed rejections among 120 author-designed
references, not evidence of a zero population false-positive rate. Family-level
estimates appear in Table~\ref{tab:family-results}; stage-specific attribution
appears in Figure~\ref{fig:constructed-stage-matrix}.

\begin{table}[t]
\caption{Recall by defect family and cumulative configuration.}
\label{tab:family-results}
\begin{tabular}{lrrr}
\toprule
Family & Mutants & P0--P2 recall & P0--P4 recall \\
\midrule
Security & \result{T-SEC-N} & \result{T-SEC-P2} & \result{T-SEC-P4} \\
Dependency & \result{T-DEP-N} & \result{T-DEP-P2} & \result{T-DEP-P4} \\
Privacy & \result{T-PRI-N} & \result{T-PRI-P2} & \result{T-PRI-P4} \\
Authority & \result{T-AUT-N} & \result{T-AUT-P2} & \result{T-AUT-P4} \\
Provenance & \result{T-PRO-N} & \result{T-PRO-P2} & \result{T-PRO-P4} \\
Lifecycle & \result{T-LIF-N} & \result{T-LIF-P2} & \result{T-LIF-P4} \\
Reliability & \result{T-REL-N} & \result{T-REL-P2} & \result{T-REL-P4} \\
Accessibility & \result{T-ACC-N} & \result{T-ACC-P2} & \result{T-ACC-P4} \\
\bottomrule
\end{tabular}
\end{table}

\subsection{Stage complementarity and ablations (RQ2)}

After preceding stages were accounted for, P3 uniquely detected
\result{R-P3-UNIQUE-N} mutants across \result{R-P3-UNIQUE-FAMILIES} families,
and P4 uniquely detected \result{R-P4-UNIQUE-N}. Removing P3 changed recall by
\result{R-ABLATE-P3}; removing P4 changed recall by \result{R-ABLATE-P4}.
Order-averaged component contributions were \result{R-SHAPLEY-SUMMARY}. These
results \result{R-RQ2-INTERPRETATION}. P4 reason attribution identified
\result{R-P4-RECORD-N} record-mediated detections and
\result{R-P4-RUNTIME-N} detections from the two executable runtime relations in
the primary constructed stratum. In the bounded-repair generated stratum, P4
found \result{R-REPAIR-P4-MODES}. A post-hoc source-level capability probe then
produced \result{R-RUNTIME-PROBE}. The primary mutant corpus therefore supports
the record-mediated component; the generated stratum and targeted probes supply
separate, explicitly non-confirmatory evidence that the runtime relations fire.

\subsection{Exploratory generated candidates (RQ1 and RQ3)}

All \result{R-GEN-RAW-N} frozen generations completed successfully at the model
inference layer. Under the primary bare-JSON rule, \result{R-GEN-PARSE-YIELD}
formed a package, so no primary candidate reached P0. The post-hoc
single-fenced-JSON normalizer recovered \result{R-GEN-FENCED-PARSE-YIELD}, but
\result{R-GEN-FUNCTIONAL-YIELD} passed P0. Its exclusive failure decomposition
was \result{R-GEN-FENCED-FAILURE-TAXONOMY}. Generation diagnostics found
\result{R-GEN-CAP-PROXY}; output truncation is therefore not a supported dominant
explanation, although the missing native finish reason prevents a definitive
claim.

The bounded source-repair policy was applicable to \result{R-REPAIR-ELIGIBLE}
and changed \result{R-REPAIR-CHANGED}. It enabled \result{R-REPAIR-P0} to pass
P0; every one of those 112 P0-passing candidates had actually been changed by
at least one allowed adapter. All \result{R-REPAIR-DEFECTIVE} were nonconforming, including
\result{R-REPAIR-MULTIFAULT}; the observed family counts were
\result{R-REPAIR-FAMILIES}. Conventional P0--P2 detected
\result{R-REPAIR-P2}, whereas P0--P4 detected \result{R-REPAIR-P4}, a
\result{R-REPAIR-DIFF} difference. This is downstream evidence on a selected,
post-hoc repair stratum, not an estimate of unassisted model reliability.

\begin{figure}[t]
  \centering
  \includegraphics[width=\linewidth]{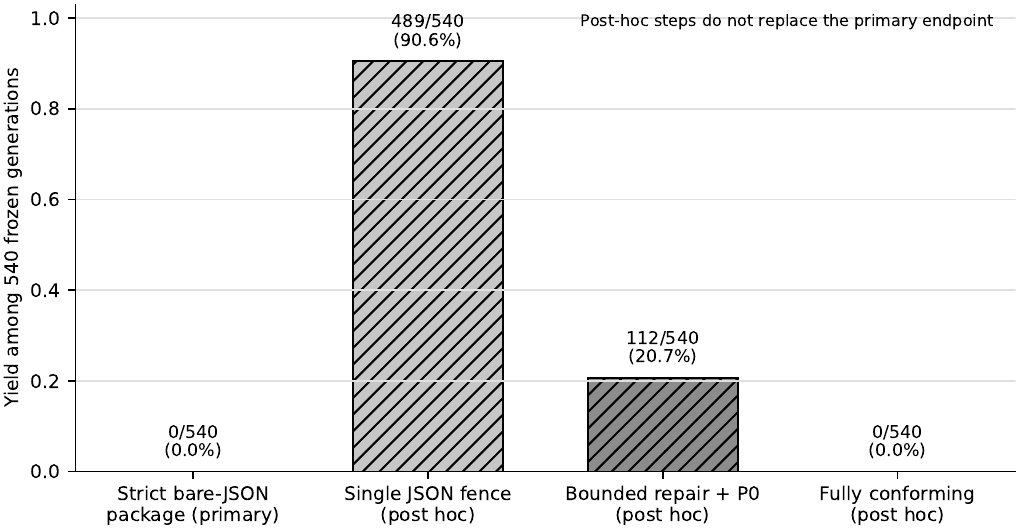}
  \caption{Primary and post-hoc generation yields. The strict bare-JSON result
  is the primary endpoint. Hatched bars are successive, explicitly post-hoc
  sensitivities and do not replace it; the final bar requires P0--P4
  conformance after bounded repair.}
  \Description{Four-bar chart. Strict bare-JSON package yield is zero of 540;
  single-fence normalization yields 489 of 540; bounded repair enables 112 of
  540 to pass P0; none of 540 is fully conforming.}
  \label{fig:generated-pipeline}
\end{figure}

Relative to the interface-only prompt, the assurance-aware prompt changed
normalized parse yield by \result{R-PROMPT-PARSE-DIFF}
(specification-clustered 95\% interval \result{R-PROMPT-PARSE-DIFF-CI}). Full
conformance remained 0/270 under each prompt; the all-zero floor does not
establish equivalence and \result{R-GEN-MODEL-FIT}. Post-repair exact-source
deduplication retained \result{R-REPAIR-DEDUP}; because \texttt{ast.unparse}
normalizes formatting and removes comments whenever an adapter fires, this is a
deduplication of the serialized repaired sources rather than the raw model text.
P0--P4 still detected every retained
P0-passing defective candidate. Clean-package harm is not estimable within this
generated stratum because no bounded-repair candidate was clean
(Figure~\ref{fig:generated-pipeline}).

\subsection{Confirmatory unmodified generation transfer (S6)}
\label{sec:confirmatory-results}

The raw-archive validator matched every identity, prompt digest, model revision,
seed, and token count against the frozen 600-row matrix. All
\result{R-CONFIRM-RAW} were present with generation status \texttt{ok}; none
reached the 4,096-token ceiling. Under the frozen parser,
\result{R-CONFIRM-PARSE}. Model-specific outcomes expose two distinct interface
failures: \result{R-CONFIRM-QWEN}, whereas \result{R-CONFIRM-DEVSTRAL}.
Consequently, \result{R-CONFIRM-P0-FULL}. The exclusive first-failure taxonomy
is therefore 300 package-parse failures and 300 visible-contract failures.

Within each model, the two prompt conditions had identical parse, P0, and
full-conformance outcomes. The assurance-aware-minus-contract-only contrasts
were \result{R-CONFIRM-PROMPT}. Because no unmodified candidate passed P0,
P0--P2 recall, P0--P4 recall, their difference, and clean-candidate rejection
are undefined in S6. Reporting them as zero would conflate absence of eligible
packages with detector failure. The confirmatory result therefore narrows the
generated-code claim to interface and visible-functional yield; it supplies no
unmodified downstream effectiveness estimate.

\subsection{Independent Moodle security transfer (S7)}
\label{sec:external-results}

Of 18 seeded announcements, \result{R-EXTERNAL-DESIGN} were retained; MSA-26-0012
and MSA-26-0015 were excluded before detector execution because each MDL search
resolved to two candidate fixing commits. The retained corpus contains 28
matched PHP file pairs. The frozen generic PHP rules detected
\result{R-EXTERNAL-VULNERABLE}, rejected \result{R-EXTERNAL-FIXED}, and produced
\result{R-EXTERNAL-DIRECTION}; exact matched discordance was
\result{R-EXTERNAL-MCNEMAR}.

This is a clear negative transfer result for the evaluated detector and corpus,
not a validation success. The post-hoc descriptive audit accounts for every
miss: five fixes add Moodle-specific session-key checks, four add
application-specific authorization constraints, and the remainder require
framework-aware identity, object-scope, resource-bound, URL-guard, path-flow,
or sanitization semantics. The frozen generic rules encode none of those
missing invariants at the required application context. The audit does not add
rules or revise the primary decisions.

\subsection{Held-out generalization (RQ3)}

On five unseen specifications, P0--P4 recall was
\result{R-HELDOUT-RECALL} and clean-package false-positive rate was
\result{R-HELDOUT-CLEAN-FPR}. The recall difference over P0--P2 was
\result{R-HELDOUT-DIFF} (95\% CI \result{R-HELDOUT-DIFF-CI}). For the
\result{R-HELDOUT-OP-N} mutants produced by the eight operators reserved for
those same specifications, the joint fixed-set difference was
\result{R-HELDOUT-OP-DIFF} (clustered resampling interval
\result{R-HELDOUT-OP-DIFF-CI}). The latter interval collapses because every
held-out specification has the same balanced operator composition and observed
difference; it is not evidence of zero uncertainty beyond this designed set.
The post-hoc factorial challenge applied those eight operators across all 30
specifications and preserved a \result{R-FACTORIAL-DIFF} advantage:
\result{R-FACTORIAL-CHALLENGE}; its balanced fixed-set
design-resampling range describes this design rather than a population of
operators. The primary, held-out, bounded-repair, and factorial contrasts use
30, 5, 28, and 30 contributing specification clusters, respectively.
Leave-one-archetype-out estimates ranged over
\result{R-ROBUSTNESS-RANGE}. The evidence therefore
\result{R-RQ3-INTERPRETATION}. In the generated held-out partition, the primary
and format-only outcomes were \result{R-GEN-HELDOUT-SUMMARY}; bounded repair
enabled \result{R-REPAIR-HELDOUT}.

\subsection{Operational governance (RQ4)}

Median end-to-end detector timing was \result{R-LATENCY-CONSTRUCTED} for the
constructed stratum. For format-normalized generated packages, the failed P0
attempts took \result{R-LATENCY-GENERATED}; no strict candidate reached P0.
The measurements are environment-specific and include the shared sandbox cost.

Calibration selected \result{R-P6-THRESHOLDS}. On held-out rows, P6 rejected
\result{R-P6-HELDOUT-CLEAN-REJECT} and quarantined
\result{R-P6-HELDOUT-CLEAN-QUARANTINE} of clean packages; its overall
quarantine rate was \result{R-P6-HELDOUT-QUARANTINE}. It detected
\result{R-P6-HELDOUT-DETECTION} of nonconforming candidates and left
\result{R-P6-HELDOUT-RESIDUAL-RISK} of severity weight admitted.
Alternate severity weights and quarantine-cost coefficients
\result{R-P6-SENSITIVITY}. Its held-out decisions were
\result{R-P6-BASELINE} reject-on-any-finding, and it never exercised
quarantine. Generated decisions are P0 fail-closed consequences. Thus P6
demonstrates transparent policy selection, not superiority over the simpler
baseline.

\begin{figure}[t]
  \centering
  \includegraphics[width=0.88\linewidth]{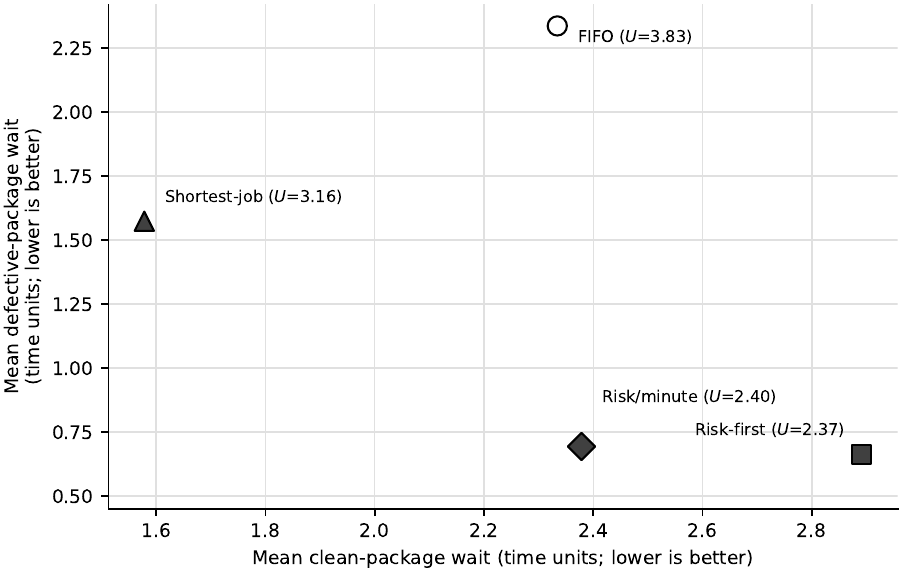}
  \caption{Paired-trace queue trade-off. Each marker reports mean
  clean-package and defective-package waiting time for one policy; $U$ in its
  label is the mean severity of defective jobs unfinished at the final-arrival
  horizon. Filled markers denote nondominated policies.}
  \Description{Scatter plot with lower waiting time preferred on both axes.
  FIFO has the longest defective-package wait. Risk-first and risk-per-minute
  have the shortest defective-package waits but longer clean-package waits.
  Shortest-job-first has the shortest clean-package wait and an intermediate
  defective-package wait. FIFO is the only unfilled, dominated marker.}
  \label{fig:queue-tradeoff}
\end{figure}

Across 10,000 paired replications, the risk-per-minute policy changed mean
defective-package waiting time by \result{R-QUEUE-DEFECT-DIFF} relative to FIFO
and changed clean-package waiting time by \result{R-QUEUE-CLEAN-DIFF}. Its
unfinished defective-job severity at the final-arrival horizon,
95th-percentile delay, and throughput were
\result{R-QUEUE-RISK}, \result{R-QUEUE-P95}, and
\result{R-QUEUE-THROUGHPUT}. The Pareto analysis identifies
\result{R-QUEUE-PARETO} (Figure~\ref{fig:queue-tradeoff}); across seven post-hoc
runs, \result{R-QUEUE-SCENARIOS}. These are parameter-grid scenarios, not
institutional measurements, and the clean-wait reversal precludes universal
dominance.

\section{Discussion}

\subsection{Assurance beyond functional correctness}

The central engineering distinction in EduPluginBench is between implementing a
function and qualifying a component for admission. Build and functional tests
answer whether a package can execute its requested path. Domain contracts answer
whether that execution remains inside declared authority, data, evidence, and
lifecycle boundaries. The observed difference,
\result{R-DISCUSSION-LAYERED-EVIDENCE}, indicates
\result{R-DISCUSSION-LAYERED-IMPLICATION}.

This distinction does not make conventional tooling less important. Security
and dependency stages cover defect mechanisms for which mature analyzers and
advisory ecosystems exist. Domain monitors add a layer whose inputs are explicit
ecosystem policy. A practical pipeline should preserve both, expose reasons from
each, and avoid presenting a single aggregate ``safety score'' as a universal
property.

\subsection{Generated-code governance}

The two generation studies answer different questions. S3 diagnoses an earlier
540-candidate protocol and, after explicitly post-hoc bounded repair, creates a
selected subset in which downstream stages can be compared. All 112 repaired
P0 passes remained nonconforming; P0--P2 detected 15, whereas P0--P4 detected
all 112. This is evidence about the selected repaired subset, not unassisted
generation. Within S3, \result{R-DISCUSSION-PROMPT-EVIDENCE}, indicating
\result{R-DISCUSSION-PROMPT-IMPLICATION}; that exploratory contrast does not
identify a general prompt effect.

S6 removes that repair dependency, increases the response ceiling, uses an
exact human-readable scaffold, and freezes every decision before inference.
Its 300 parse failures and 300 visible-contract failures still yield no
unmodified P0 pass. The model-specific split is informative: one model omitted
required package declarations, while the other wrapped every response in an
unlabelled fence excluded by the frozen interface. Neither outcome estimates
P1--P4 effectiveness. Together, S3 and S6 show that interface compliance,
visible functionality, and release-contract conformance are separate gates;
they do not establish that the full pipeline detects defects in unmodified,
functionally valid model-generated plugins.

\subsection{Independent detector transfer}

S7 provides construction-independent labels for a narrow security-facing test,
and its result is negative. Generic PHP Semgrep rules did not flag any of the 16
vulnerable Moodle revisions or distinguish them from their fixes. The public
diff audit shows why: the fixes encode absent framework guards, capability and
ownership constraints, cross-file session-key propagation, and other Moodle
semantics rather than generic syntax patterns. This result strengthens the
paper's calibration by falsifying a broad transfer reading. Generic SAST remains
useful for the vulnerability mechanisms it models, but governed ecosystems need
application-aware contracts and independent real-defect validation of those
contracts.

\subsection{Implications for plugin-platform engineers}

The results motivate four deployment practices, conditioned on the reported
evidence.
\begin{enumerate}
  \item Express role, telemetry, provenance, lifecycle, and failure obligations
  as versioned, executable contracts rather than prose review criteria.
  \item Preserve detector reasons and normalized traces so that quarantine and
  repair decisions are auditable.
  \item If a platform uses graded scores or quarantine, calibrate them on
  clean-package harm and residual severity-weighted risk, not recall alone.
  \item Use operational simulation to choose review priorities, and publish the
  scenario assumptions that drive the selected policy.
\end{enumerate}
The extent to which each recommendation is supported is summarized by
\result{R-PRACTICE-SUPPORT-SUMMARY}. In this corpus, calibration produced no
held-out decision advantage over reject-on-any-finding and did not exercise
quarantine; the third recommendation is consequently conditional design advice,
not an observed benefit.

\subsection{Generalization beyond education}

Education supplies concrete role, evidence, telemetry, and state semantics, but
the method is not tied to a learning-outcome claim. Other governed plugin
ecosystems---for example, enterprise workflow, scientific analysis, or public
administration---also combine conventional software properties with
domain-specific authority and provenance. Transfer requires a new specification
schema, clean corpus, and domain oracles; the present effect sizes should not be
carried over without that revalidation. The Moodle result further shows that a
generic language ruleset is not a substitute for ecosystem semantics.

\subsection{Porting recipe}

Porting EduPluginBench to another governed plugin ecosystem requires eight
auditable steps:
\begin{enumerate}
  \item define the package schema, threat model, and release-blocking contract;
  \item obtain independently authored clean references where feasible;
  \item author fault transformations and separate construction oracles, then
  document every field shared with evaluated detectors;
  \item freeze splits, tool versions, rule sets, prompts, seeds, and endpoints;
  \item validate fault activation, isolation, and immutable parent lineage;
  \item evaluate unmodified model-generated packages without replacing failed
  rows through outcome-dependent repair;
  \item acquire independently labelled vulnerable/fixed revisions and test the
  relevant conventional and domain detectors without relabelling; and
  \item calibrate admission on clean harm and residual risk, then reproduce all
  claims from row-level evidence and checksums.
\end{enumerate}
The present artifact implements each step, but independent authorship of more
reference and oracle material remains a priority for replication.

\section{Threats to Validity}
\label{sec:threats}

\subsection{Construct validity}

Mutation score is not defect prevalence. First-order mutants isolate stage
effects but may be easier to detect than naturally occurring multi-fault code.
Activation and isolation checks, explicit parent lineage, generated candidates, and held-out
operators mitigate this threat without eliminating it. The eight families are a
declared operationalization of release risk, not an exhaustive definition of
plugin safety.

The bounded-repair stratum is post hoc and selected by syntactic validity plus
three visible-contract transformations. It enables downstream measurement but
cannot estimate unassisted generation yield or defect prevalence among all model
outputs. The policy was frozen before its run, excluded hidden feedback and
metadata changes, and preserved the strict endpoint. Python AST serialization
normalizes formatting and removes comments when a repair fires; post-repair
exact-source deduplication and held-out-specification summaries expose sensitivity within
the selected stratum; replication on unmodified, P0-passing model outputs
remains necessary. S6 performs that unmodified replication but produces no P0
pass. Its model-specific parse split partly measures compliance with the exact
response interface: an unlabelled outer fence is a reproducible integration
failure under the frozen contract, not evidence that the enclosed source is
semantically incorrect. The study therefore supports end-to-end package yield,
not an isolated code-quality ranking.

Construction oracles and evaluated detectors share a specification vocabulary.
They are implemented as separate functions; detector code does not inspect
labels, and stripping the lineage fields leaves decisions unchanged. The package
object may still carry those fields, and several P3/P4 decisions and their
corresponding reference oracles inspect
the same manifest or behavioral fields. High controlled-mutant recall therefore
shows consistency with the encoded contract; it does not by itself establish
independent discovery of real defects. Activation and isolation checks,
containerized metamorphic relations, generated multi-fault candidates, held-out
operators, factorial operator application, source-level runtime probes,
negative cases, and quarantined ambiguity reduce this threat without removing
it. The runtime probes are capability tests deliberately targeted to the two
relations, not unbiased effectiveness estimates. A replication with
independently authored implementations,
behaviorally richer APIs, and black-box outcome oracles is needed.

The 120 clean packages are instantiations of four author-designed source
families, not 120 independently authored systems. Their variation supports
false-rejection and regression checks across multiple structures but cannot
represent the diversity of production plugins. We therefore report the Wilson
upper bound when no clean package is rejected and avoid describing zero
observed rejections as proof of zero false-positive risk.

The external Moodle labels are independent of the detector, but S7 evaluates
only one frozen generic PHP ruleset over changed files. A zero finding can
reflect absent framework semantics, limited interfile analysis, or corpus
packaging rather than the absence of detectable security signals in principle.
The paired revisions and complete miss audit make that failure inspectable; they
do not validate EduPluginBench's six non-security families or a hypothetical
Moodle-specific detector.

\subsection{Internal validity}

Tool configuration, advisory drift, sandbox failures, and accidental split
leakage can bias results. The study pins detector versions and rules, archives
the advisory snapshot, records infrastructure failures as abstentions after one
controlled rerun, partitions by specification, and scans canonical digests and
token fingerprints for duplicates. Generation and pipeline timings remain
environment-specific; platform, accelerator, and tool metadata are retained,
and timing outcomes are interpreted descriptively rather than as portable
performance estimates. The confirmatory raw archives were validated against all
600 frozen matrix identities before parsing. Moodle source bytes, labels,
parent/fix revisions, and the Semgrep ruleset were separately locked before
detector execution; two ambiguous histories were excluded under the frozen rule
rather than resolved after seeing findings.

\subsection{Conclusion validity}

Mutants from one specification are dependent, so row-level binomial intervals
would be overconfident. The specification is the bootstrap cluster and random
intercept. Predeclared secondary tests use Holm correction, and interpretation
emphasizes absolute effects and clustered uncertainty rather than the
row-level McNemar probability. With only five held-out
specifications, family-specific held-out intervals may remain wide; the paper
reports them rather than substituting calibration performance. Because the
predeclared held-out operators occur only on those specifications, that result
cannot identify separate operator and specification contributions. The
post-hoc factorial challenge removes this confound for the fixed eight
operators but does not justify inference to an operator population; its
degenerate clustered interval reflects the exactly balanced design.

\subsection{External validity}

The constructed corpus covers Python packages and five educational archetypes.
The two generation studies cover five downloadable coding-model families across
two protocols and one Colab execution environment; outcomes reflect four-bit
quantization and their respective 1,600- and 4,096-token ceilings. S7 adds one
public PHP application and one generic detector configuration. Different
languages, proprietary frontier models, larger repositories,
production dependency graphs, and institution-specific policies may yield
different results. The selected models are constrained by reproducibility,
licensing, and standard Colab hardware; they are not claimed to represent all
current code generation. The three-rule bounded repair is specific to this
visible interface and is not proposed as a general repair algorithm.
Scenario-based review times, including the seven-run sensitivity grid, do
not estimate the behavior of real reviewers or institutions.

\section{Ethics, Safety, and Reproducibility}

\subsection{Human-subject boundary}

The study recruits no participants, observes no developers, learners, teachers,
or reviewers, and processes no personal or institutional records. All
identities, course events, submissions, and evidence are synthetic. The study
therefore makes no empirical claim about learning, usability, trust, human
review quality, or institutional operating performance. S7 uses only public
security announcements and public source-code revisions; these are software
artifacts, not participant data.

\subsection{Safe handling of generated and vulnerable code}

Model-generated code and deliberate vulnerable fixtures are treated as untrusted.
They are never deployed, receive no production credential, and run only after a
pre-execution gate in a network-isolated, resource-limited ephemeral sandbox.
Historical Moodle PHP is never executed and is scanned only as inert text.
The public artifact marks vulnerable fixtures conspicuously and defaults to
non-execution. Model licenses and generated-code attribution risks are recorded;
raw candidates are distributed only when their applicable terms permit it.

\subsection{Artifact organization}

The research compendium contains:
\begin{itemize}
  \item frozen protocols and amendments;
  \item executable specifications, clean-package generators, and conformance checks;
  \item mutation registry, exact operator definitions, parent lineage,
  activation results, and invalid attempts;
  \item exact model revisions, prompts, seeds, raw generations, and extraction logs;
  \item both immutable 300-row confirmatory model archives, the complete
  600-row decision table, and the prespecified analysis;
  \item public Moodle announcement and revision provenance, vulnerable/fixed
  file pairs, frozen Semgrep output, exclusions, and the 16-row miss audit;
  \item detector versions, rules, SBOMs, advisory snapshot, and sandbox configuration;
  \item row-level observations, exclusions, traces, and timing data;
  \item deterministic analysis, table, and figure generators;
  \item the result-value and claim-to-evidence registries;
  \item checksums, software bill of materials, smoke tests, and clean-room instructions.
\end{itemize}
The clean-room verification workflow verifies checksums, asserts the protocol
lock and installed tool versions, runs automated and container checks,
recomputes analysis summaries and manuscript values from frozen rows,
regenerates figures, and recompiles the article.

\section{Conclusion}

EduPluginBench frames admission of model-generated plugins as a
software-assurance problem spanning code, dependencies, runtime behavior, and
domain policy. On the constructed benchmark, the full pipeline
\result{R-CONCLUSION-EFFECT}; this establishes consistency with the encoded
contract, not independent production-defect effectiveness. The transfer studies
then delimit that result. In S6, \result{R-CONFIRM-PARSE} parsed but
\result{R-CONFIRM-P0-FULL}, leaving downstream assurance effects undefined. In
S7, \result{R-EXTERNAL-RESULTS}, demonstrating that a generic language ruleset
did not capture the evaluated application's security invariants. The earlier
bounded-repair diagnostic and held-out operator challenges
\result{R-CONCLUSION-GENERALIZATION}; they remain supporting, explicitly
selected or fixed-set evidence. Queue results remain scenario-only.
Accordingly, the paper supports \result{R-CONCLUSION-CLAIM}, while rejecting a
claim of universally safe plugins or proven real-defect transfer. The
checksummed compendium enables independent inspection of every retained package,
public-source pair, decision, exclusion, and reported value.

\begin{acks}
OpenAI Prism and Codex supported implementation scaffolding, manuscript editing,
and consistency checks under the verification boundaries reported in
Section~\ref{sec:reproducibility-controls}. Experimental model outputs supplied
neither oracle labels nor detector decisions. Nizam Kadir reviewed every change,
is the sole human author, and accepts responsibility for the originality,
accuracy, and integrity of the work.
\end{acks}

\bibliographystyle{ACM-Reference-Format}
\bibliography{references}

\end{document}